\documentclass{rmaa} % for a referee version
\usepackage{graphicx}
\usepackage{natbib}
\input{epsf.sty}

\def\nh{$n_{\rm H}$\/}
\def\mbh{$M_{\rm BH}$\/}\def\rfe{R$_{\rm FeII}$}
\def\feiiq{\rm Fe{\sc ii }$\lambda$4570\/}

\def\msol{M$_\odot$\/}

\def\ltsima{$\; \buildrel < \over \sim \;$}
\def\simlt{\lower.5ex\hbox{\ltsima}}            % < over MMM
\def\gtsima{$\; \buildrel > \over \sim \;$}

\def\simgt{\lower.5ex\hbox{\gtsima}}            % > over MMM

\def\ha{{\sc H}$\alpha$}

\def\civ{{\sc{Civ}}$\lambda$1549\/}

\def\civbc{{\sc{Civ}}$\lambda$1549$_{\rm BC}$\/}

\def\cm3{cm$^{-3}$\/}
\def\hb{{\sc{H}}$\beta$\/}

\def\hbbc{{\sc{H}}$\beta_{\rm BC}$\/}

\def\mgii{{Mg\sc{ii}}$\lambda$2800\/}

\def\ciii{{\sc{Ciii]}}$\lambda$1909\/}
\def\oiiiopt{{\sc{[Oiii]}}$\lambda$4959,5007\/}
\def\o4363{{\sc{[Oiii]}}$\lambda$4363\/}

\def\siiii{{Si}{\sc iii}]$\lambda$1892\/}

\def\feiiuv{{{Fe{\sc ii}}}$_{\rm UV}$\/}
\def\feiiopt{{Fe\sc{ii}}$_{\rm opt}$\/}
\def\feii{{Fe\sc{ii}}\/}
\def\fe{{\sc{Fe}}\/}
\def\dvr{{$\Delta v_{\mathrm r}$}}
\def\vr{{$v_{\mathrm r}$}}

\def\fe76087{{\sc [Fe vii]}$\lambda$6087\/}
\def\oiii{{\sc [Oiii]}$\lambda$5007}

\def\kms{km~s$^{-1}$}

\def\gs{$\Gamma_{\rm soft}$\/}
\def\lm{$L_{\rm bol}/M_{\rm BH}$}

   \title{Low Redshift BAL QSOs in the Eigenvector 1 Context}

   \author{    J. W. Sulentic\altaffilmark{1},
          D. Dultzin-Hacyan\altaffilmark{2},
          P. Marziani\altaffilmark{3},
           C. Bongardo\altaffilmark{3},
           V. Braito\altaffilmark{3},
          M. Calvani\altaffilmark{3}, \&
            R. Zamanov\altaffilmark{3,4}
   \altaffiltext{1}{Department of Physics and Astronomy, University of
              Alabama, USA}
         \altaffiltext{2}    {Instituto de Astronom\'\i a, UNAM,  M\'exico}
      \altaffiltext{3}{INAF, Osservatorio Astronomico di Padova, Italy}
            \altaffiltext{4}{National Astronomical Observatory, Bulgaria}
   }
\shortauthor{Sulentic et al.}
\shorttitle{Low-$z$ BAL AGNs} \fulladdresses{ \item Jack W.
Sulentic: Department of Physics and Astronomy, University of
Alabama, Tuscaloosa, AL 35487, USA; e-mail:
              {\tt giacomo@merlot.astr.ua.edu} \item Deborah Dultzin-Hacyan: Instituto de
Astronom\'\i a, UNAM, Aptdo. Postal 70-264, M\'exico, D.  F.
04510, M\'exico;  e-mail: {\tt deborah@astroscu.unam.mx} \item
Paola Marziani, Christian Bongardo, Valentina Braito, \& Massimo
Calvani: Istituto Nazionale di Astrofisica, Osservatorio
Astronomico di Padova,  Vicolo dell' Osservatorio 5, I-35122
Padova, Italy;  email: {\tt marziani@pd.astro.it;
zamanov@pd.astro.it; bongardo@pd.astro.it;
calvani@pd.astro.it}\item Radoslav K. Zamanov: National
Astronomical Observatory Rozhen, POB 136, Smoljan, BG-4700,
Bulgaria }

\resumen{}

\abstract{We attempt to characterize the geometry of Broad Absorption Line (BAL) QSOs by studying a low redshift
sample of 12 sources. We find that the majority of these sources are Population A quasars as defined in \citet[][broad
\hb\ FWHM$\leq$ 4000 \kms]{sulenticetal00a}. A possible correlation between terminal velocity and absolute $V$\
magnitude  suggests that the bolometric luminosity to black hole mass ratio \lm\ is a governing factor with classical
BAL sources showing the highest values. \civ\ emission in classical BAL sources  shows a profile blueshift that
supports a disk wind/outflow scenario with a half opening angle of $\la$ 50$^\circ$. Observation of ``secondary''
mini-BAL features in the \civ\ emission profile motivates  us to model BALs with an additional component that may be
involved with  the BLR outflow and  co-axial with the accretion disk.}

\addkeyword{Quasars} \addkeyword{Galaxies:
Active}\addkeyword{Lines: Absorption} \addkeyword{Lines: Emission}

\begin{document}
\maketitle

\section{Introduction \label{intro}}

Type 1 Active Galactic Nuclei (AGNs) apparently show three kinds
of associated absorption features (here we focus on \civ\  as a
characteristic high ionization line, HIL).

\begin{enumerate} \item Narrow absorption
lines (NALs) are the most common and are accepted as
``associated'' \citep[e. g.,][]{vestergaard03} (1) if they lie
within $\pm$5000 \kms\ of the \civ\ emission line centroid, (2) if
the rest frame equivalent width is between 0.5 and $\approx$ 4.0
\AA, (3) if the line width is no more than a few 100 \kms (or
alternatively the \civ\ doublet is resolved in absorption), and/or
(4) if the lines are variable in intensity
\citep{richardsetal99,richardsetal01,brandtetal00,gangulyetal01,vestergaard03,wiseetal04}.
%More controversial are possibly associated
%NALs that are blueshifted in the range 5000-25000 \kms. Despite
%contradictory claims, NALs do not appear to occur preferentially in
%any particular class of Type 1 AGNs.

\item Broad absorption lines (BALs) are thought to occur in 15 \%\
-- 20 \%\ of AGNs \citep{hewettfoltz03,reichardetal03} and
generally show blueshifted troughs with velocity widths of several
thousands \kms\ and equivalent widths of several tens of \AA.
Terminal velocities in these broad features are measured in tens
of thousand \kms.

\item Absorption features with intermediate properties (FWHM~few
thousand \kms, EW $\sim 5-10$ \AA\ and sometimes variable) are
called ``mini-BALs'' and may be even rarer than BALs
\citep{jannuzietal98,narayanetal04}.
\end{enumerate}

It is tempting to relate mini-BALs to BALs because they show
blueshift velocities up to ~50000 \kms, overlapping the BAL range,
although there are arguments against a close connection \citep[e.
g.,][]{narayanetal04}. However, mini-BALs are regarded as
intrinsic (i.e., due to gas physically close to, and dynamically
affected by the active nucleus central engine) absorption
features, and not just as ``associated'' like NALs, which could be
also due to distant gas). In this paper we consider BAL and
mini-BAL phenomenology in low redshift quasars, while only
occasional consideration will be given to NALs.
%So far few BALs or
%mini-BALs have been discovered in radio-loud (RL) sources and some
%claimed RL BALs (e.g. Becker et al. 1997) turn out to be radio-quiet
%(RQ) after the optical/UV flux is corrected for internal extinction
%(Najita et al. 2000).

The discovery of BAL and mini-BAL QSOs has always been biased
toward AGNs with high enough redshift to bring the most prominent
UV resonance lines (e.g., \civ, \mgii) into the visible spectrum.
IUE ($\sim$1978) and HST ($\sim$1990) made it possible to observe
the UV rest frame in low$-z$\ AGNs resulting in the identification
of several low $z$\ sources showing BALs and mini-BALs
\citep{turnsheketal88,borosonmeyers92,barlowetal97,jannuzietal98}.

One advantage of low redshift sources involves the possibility of
a reliable determination of the quasar rest frame from optical
emission lines or, in a few cases, host galaxy measures. An
accuracy $\Delta z \la$0.001 can be achieved from \oiii\ or the
peak of the narrow component of \hb.

Another advantage stems from our ability to interpret the sources in the context of the Eigenvector 1 (E1) parameter
space \citep{sulenticetal00b}.  E1 tells us that AGN optical and UV emission line phenomenology is not randomly
dispersed around an ``average'' spectrum. We identified four parameters that provide optimal discrimination between
the various classes of broad line emitting AGNs. The three emission line parameters are : (1) FWHM of the \hb\ broad
component (\hbbc), (2) \rfe = W(\feiiq)/W(\hbbc) and (3) C(1/2) = profile centroid displacement of \civ\ at half
maximum.  They are supplemented by (4) \gs, the soft X-ray photon index. If we think of E1 as an H-R diagram for
quasars then we find a principal occupation or main sequence for the majority of low redshift sources. The sequence
ranges from radio-quiet (RQ) narrow line Seyfert 1 (NLSy1) sources with narrowest FWHM(\hbbc), strongest \rfe, largest
\civ\ blueshift and strongest soft X-ray excess to lobe-dominated radio-loud (RL) sources with the broadest \hbbc\
profiles, weakest \rfe\ and no soft X-ray excess or \civ\ blueshift.  The main E1 occupation/correlation sequence is
likely to be driven by the Eddington ratio \citep{marzianietal01,boroson02,zamanovmarziani02,yuanwills03}. The soft
X-ray excess parameter has also been interpreted in the same scenario with the sources accreting at highest \lm\
showing the largest values of $\Gamma_{\rm soft}$\ \citep{poundsetal95}.

Another way to emphasize phenomenological difference in the E1
parameter space is through the Population A--B concept. Population
A is an almost pure RQ population (with NLSy1 as the extremum)
defined by the criterion FWHM(\hbbc)$\simlt$4000 \kms\ \citep[at
low $z$; see][]{sulenticetal04}. Population B is a mixed RL and RQ
population with broader \hbbc. Type 1 AGN show some evidence for a
discontinuity between Pop. A and B in \hbbc, \civbc\ profiles and
occurrence of HIL blueshifts \citep{marzianietal03a,bachevetal04}.
Analysis of the optical E1 parameters support the idea that Pop. A
sources are radiating at higher Eddington ratio than Pop. B
\citep{borosongreen92,poundsetal95, marzianietal03b}. Further
support comes from \civ\ and other UV HIL resonance lines which
show evidence for a wind or outflow in Pop. A sources.

%that might be powered by high accretion.

%Low redshift BAL and mini-BAL may suffer from small numbers but
%they benefit from more independent pieces of information.  \S 2
%defines our BAL + mini-BAL sample followed in \S 3 by a
%description of the data analysis. In \S 4 and \S 5, respectively,
%discuss the data and present a simple model.

\section{Sample Definition \label{sample}}
%so one can only attempt to maximize the sample.

Defining a large, complete sample of low-$z$\ BAL QSOs is
impossible with present data. We select BAL  + mini-BAL
\citep[BALnicity index $\approx$ 0 \kms;][]{weymanetal91} QSOs
using the following two criteria: (1) FWHM(\civ)$_{\rm abs} \ga$
2000 \kms, (2) and W(\civ)$_{\rm abs} \ga $ 4 \AA. This results in
negligible overlap with high and low $z$\ NAL samples
\citep[e.g.][]{richardsetal99,brandtetal00,gangulyetal01,vestergaard03}.
We consider objects with magnitude $m_{\rm V} \la 16.3 $\ and
redshift $z \la$ 0.5 and find 12 sources which include:  (a) AGNs
identified in NED as BAL QSOs \citep[with the exception of PG
1416$-$129 which does not show absorption in HST spectra;
see][]{greenetal97}, (b) BAL QSOs in \citet{turnsheketal97}
selected on the basis of weak \oiiiopt\ emission (4 objects), (c)
soft X-ray weak PG QSOs with W(\civ) $\simgt$ 5 \AA\ from
\citet{brandtetal00}, with the exception of PG 1126$-$041. We also
visually inspected the HST-FOS archival spectra that included
\civ\ \citep[as retrieved by][]{bachevetal04} and found no
additional sources satisfying our selection criteria.

Our conditions are less restrictive than those of
\citet{weymanetal91} since we consider BAL sources  with shallow
absorption troughs and with FWHM(\civ)$_{\rm abs} \ga$ 2000 \kms:
\citet{weymanetal91} exclude source with absorption trough $\la
10$\% of the adjacent continuum. We do not consider as BAL quasars
sources with absorption of any depth but with width $\la 2000$
\kms\  \citep[notable examples include Akn 564 and IRAS
04505--2958;][]{bechtoldetal02,crenshawetal02}.

Table \ref{tab:optobs} identifies the objects in our sample
(Column 1: IAU code; Column 2:  common name), and provides the
apparent visual magnitude (Col. 3; the objects reported with two
decimal digits are new observations; see \S \ref{oo}), the
absolute V magnitude (assuming $H_0 = 75$~ \kms Mpc$^{-1}$,
$\Omega_{\mathrm M}= 0.3$, and $\Omega_{\mathrm \Lambda}= 0.7$;
Col. 5), the heliocentric redshift and its associated uncertainty
(Col. 4; see also the beginning of \S \ref{results}), the date of
optical spectroscopic observations (Col. 6), the universal time at
the beginning of the exposure (Col. 7) and the exposure time in
seconds (Col. 8). The last column (Col. 9) provides the acronym of
the observatory where the data have been collected. We  use the
term ``low-$z$\ QSOs'' for all of our objects even if a few
objects are at or below the formal boundary between Seyferts and
QSOs ($M_{\rm B} \approx$ -22.1).

\section{Observations \& Data Analysis \label{analysis}}

\subsection{Observations \label{obs}}

\paragraph{Optical Photometric Observations} $B$, $V$, $Rc$, $Ic$\   observations
have been obtained (16 and 17/02/2001) with the 2.0m RC telescope
of the Bulgarian National Astronomical Observatory ``Rozhen",
equipped with a Photometrics 1024$\times$1024 CCD camera.
\citet{landolt92} standards were observed before and after every
source. Apparent $V$\ magnitudes  are given in Table
\ref{tab:optobs}. The error of our magnitudes is $\pm$0.05 mag in
all bands.

\paragraph{Optical Spectroscopic Observations \label{oo}} Data used
in the present study are part of a large sample  of spectroscopic
observations  covering the \hb\ spectral region
\citep{marzianietal03c}. The dataset covers  215 AGNs with
resolution typically around 4\AA\ FWHM and S/N (in the continuum
near \hb) usually $\ga$ 20. Data on individual observations used
in this paper are reported in Table \ref{tab:optobs}.

\paragraph{UV Spectroscopic Observations} UV observations are HST/FOS with high resolution grating (yielding a precision of
$\approx$200 \kms), and STIS medium-resolution spectra covering \civ. FOS data were re-calibrated with the latest
STPOA processing scripts using IRAF/SDAS. This takes into account systematic errors in the wavelength scale of the
FOS/BL data which were not properly considered until recently \citep[e.g.,][]{kerberrosa00}. Several IUE observations
were also included with correspondingly lower precision $\Delta v_r \sim 1000$\kms. In the case of Mrk 231 IUE data
were included because only a G150L HST/FOS observation was usable. Table \ref{tab:uvobs} provides a log of the UV
archival observations employed in the present paper. Col. 2 lists the telescope and the camera; Col. 3 the grating.
Col. 4, 5, 6 provide date, universal time and exposure time respectively. Col. 7 identifies the data-set.

\subsection{Optical and UV Data Analysis: \feiiopt\ and \feiiuv\
emission \label{meas}}

Optical and UV observations were analyzed following \citet[][\hb\ and \civ]{marzianietal96,marzianietal03a} and
\citet{sulenticetal00a}.  The wavelength and flux-calibrated spectra were shifted to the rest frame and then continuum
and \feii\ subtracted.  \feii\ subtraction used a scaled and broadened template based on observations of \feii\ strong
I Zw 1. It is now possible to define a better template for both optical and UV \feii\ emission (between 1400 \AA\ and
1800 \AA). The optical one is almost identical to the one used in \citet{borosongreen92} but is based on a higher
resolution spectrum. Our \feiiuv\ template comes from two post-COSTAR I Zw 1 observations, the first covering the
G130H spectral range (dataset Y26C0204T - 13-Feb-1994) and the second covering the G190H range (dataset Y26C0103T -
14-Sep-1994). The template agrees with a theoretical model computed for hydrogen density \nh = 10$^{12}$ \cm3, and
ionizing photon density $log\Phi_{\rm H}$=17.5 cm$^{-2}$ s$^{-1}$\ \citep{verneretal99}. This may be the most
appropriate case for the overall low ionization level observed in the spectra of NLSy1s like I Zw 1
\citep{marzianietal01}. It also agrees satisfactorily with the template of \citet{vestergaardwilkes01}.

The continuum level was defined in spectral regions at $\approx$ 1450 \AA\ and 1750 \AA, where no emission features
were appreciable. For  three sources with strong \feiiuv\ emission (PG 0043+038, IRAS 07589+6508, and PG 1700+518) the
final continuum level was set after \feiiuv\ subtraction.

\section{Results }\label{results}
\subsection{\hb\ and \civ\ Profile Analysis}

\begin{figure*}
  \includegraphics[width=17.6cm, height=17.6cm, angle=0]{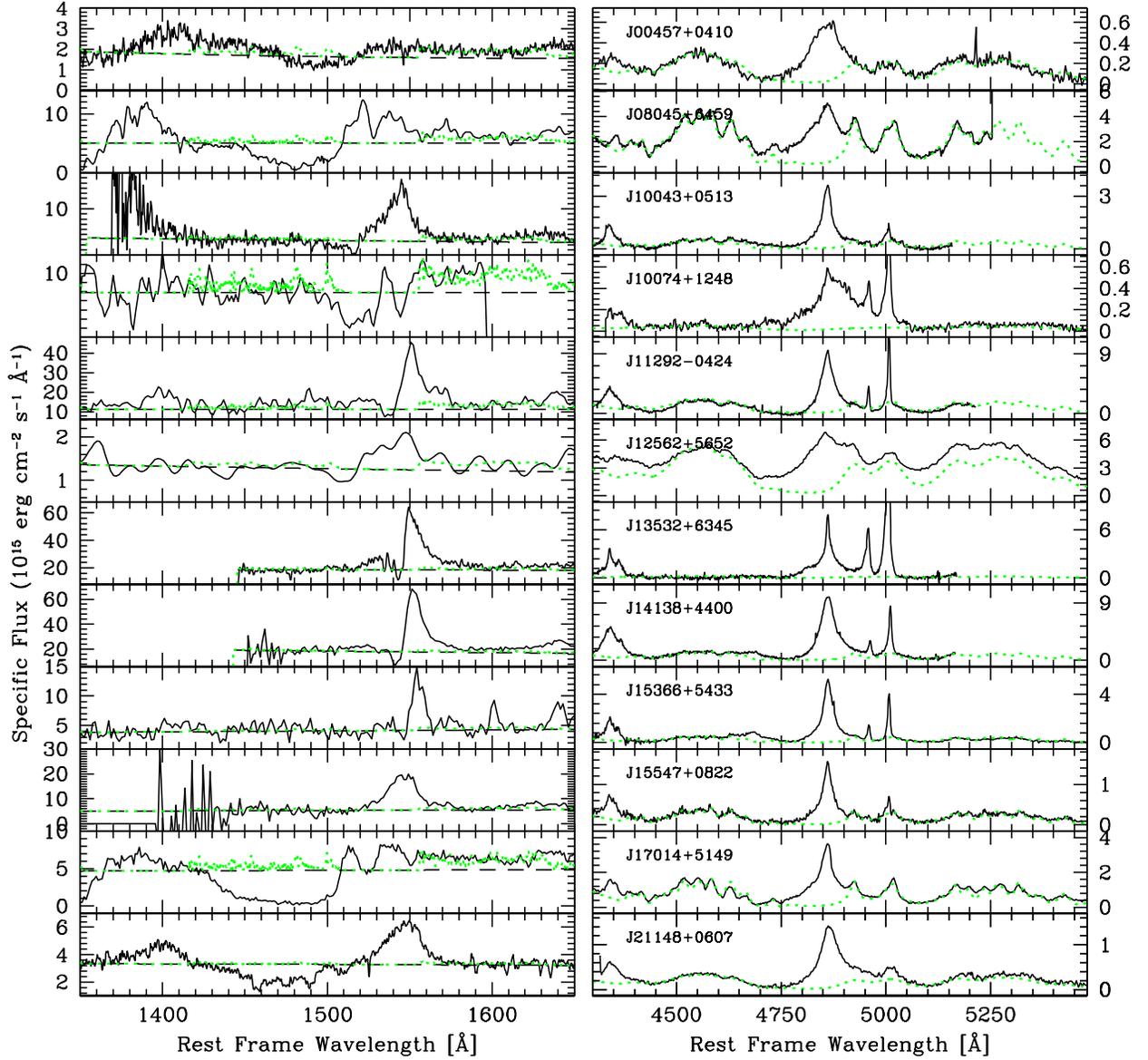}
  \caption{Deredshifted spectra for low redshift BAL QSOs: \civ\
  (left)
  and  \hb\ spectral regions (right). Solid line is the spectrum
  converted to rest frame specific flux (ergs s$^{-1}$ cm$^{-2}$ \AA$^{-1}$).
  Green dotted line indicated \feiiuv\ and \feiiopt\ emission.
  Dot-dashed lines show the continuum level.
  \label{fig:uvopt}}
 %Dot-short dash line for Mark 486 is the HST spectrum.
\end{figure*}

Rest-frame \civ\ and \hb\ spectral regions are shown in Fig. \ref{fig:uvopt} for the entire sample. The rest frame was
assumed to be equivalent to the redshift of the {\em narrow} optical emission lines except for one source where an HI
measure was available (see notes to Table \ref{oo} for references; if no reference is given then $z$\ determination
was estimated in this study).   We found no evidence for a significant discrepancy between \oiiiopt\ measurements and
other narrow lines (including the peak of \hb\ which was also used as an \oiii\ surrogate when \oiii\  was not
detected; \S \ref{e1}).

\begin{figure*}
\includegraphics[width=17.6cm, height=17.6cm, angle=0]{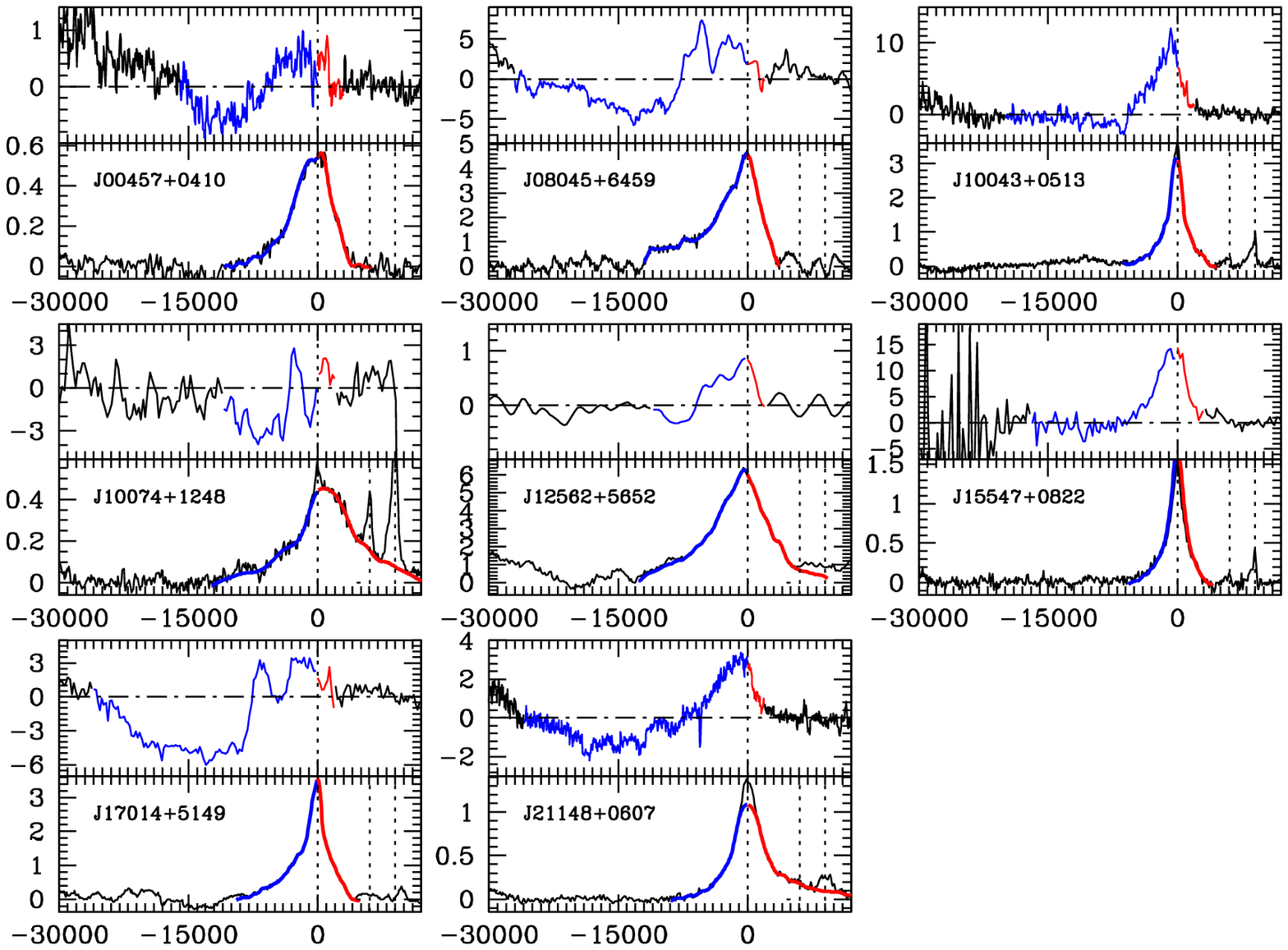}
\caption[]{Continuum and \feii\ subtracted profiles of \civ\ and
\hb\ for  8  BAL QSOs in our sample. Ordinate is rest frame
specific flux (10$^{-15}$\ ergs s$^{-1}$ cm$^{-2}$ \AA$^{-1}$) and
abscissa is radial velocity (\kms) with respect to rest frame. The
upper and lower halves of each panel  show \civ\ and \hb\ profiles
respectively.  A high order spline fit (thick line) superimposed
on the \hb\ line indicates the broad component. Dotted lines mark
the quasar rest frame  for \civ\ and \hb, as well as the expected
position of \oiiiopt. The side of the absorption/emission profiles
below and above  \vr\ $\approx$ 0 \kms\ are colored blue and red
respectively. \label{fig:civhba}
  }
 \end{figure*}

Cleaned \civbc\ and \hbbc\ profiles are shown in Fig. \ref{fig:civhba} and Fig. \ref{fig:civhbb}.  The \hbbc\ profile
is emphasized by a high-order spline following \citet{marzianietal03c}. Table \ref{tab:hb} lists the measured rest
frame parameters for the \hb\ spectral region: W(\hbbc), \rfe, FWHM(\hbbc) and FWHM(\feiiq)\ with uncertainty at a
2$\sigma$ confidence level.  Col. 2--4 of Table \ref{tab:civ} list the parameters for the emission component of \civ:
peak \vr, equivalent width, and FWHM(\civ) after \feiiuv\ correction, respectively. In addition we report: radial
velocity at minimum trough (Col 5),  maximum blueshift velocity of the absorption (i.e., the terminal velocity, Col.
6),  the equivalent width (Col. 7),  FWHM of the principal broad absorption component (Col. 8), and  BALnicity index
as defined by \citet[][Col. 9]{weymanetal91}. All values are for the source rest frame. Errors in EW and FWHM are
usually $\approx$ 20 \%. Radial velocity measurements have a similar uncertainty, unless noted. BALnicity index
uncertainty is estimated to be $\pm$ 500 \kms\ at a $2 \sigma$\ confidence level. Some values with extremely
asymmetric error bars have uncertainties indicated in the Table.

Fig. \ref{fig:civhba} shows the continuum and \feii\ subtracted \civ\ and \hb\ profiles for  BALs while Fig.
\ref{fig:civhbb} shows the spectra of the 4 mini-BALs. Fig. \ref{fig:civhbb} shows the assumed unabsorbed profile
estimated by reflecting the red unabsorbed side of the emission profile onto the blue absorbed side. This procedure
works especially well for PG 1351+640 and PG 1411+442 where the  mini-BAL troughs appear to be displaced blueward from
the peak emission. In the case of BALs  we measured the absorption equivalent width by interpolating the continuum
across the absorption profile.

Several characteristics are important in the context of BAL QSOs:

\begin{enumerate}

\item the emission component of \civ\ is almost fully blueshifted relative to the rest frame velocity
\citep[e.g.][]{reichardetal03}.

\item three out of six sources (IRAS07598+6508, PKS 1004+13, PG 1700+518) with BALnicity Index $\ga$ 0 \kms\  show a
secondary absorption trough with variable velocity (\vr) relative to the rest from source to source, but located close
to the average radial velocity of the emission component. Secondary absorptions are not easily attributed  to a
line-locking interpretation. The absorption velocity may correlate with the terminal velocity of the BAL: two sources
with $v_{\rm r, T} \sim 27000$ \kms\ have $v_{\rm r, sec} \approx 4000$\kms, while if $v_{\rm r, T} \sim 15000$ \kms,
$v_{\rm r, sec} \sim 1000$ \kms. The secondary component is apparently not rare, or restricted to our sources.
Approximately 30 \%\ of the \civ\ absorption/emission profiles in \citet{koristaetal93} include a secondary absorption
and appear similar to ours. About $50$\%\ of sources show a complex absorption pattern or a PHL~5200-like profile,
where  the \civ\ emission component appears as a spiky feature because it is almost completely eaten away by deep and
very broad absorption \citep{turnsheketal88}. This kind of feature does not seem to be present in the remaining
$\approx$20\%\ of BAL sources.

\item the depth of the associated secondary absorption component
suggests a covering factor $0.5 \la f_{\rm c}\la 1$.

\item BAL absorption apparently affects only a minor part of the emission component since spectral type A3 (the most
extreme pop A/NLSy1 subtype in  \citet{sulenticetal02} -- see Fig. \ref{fig:e1bal} in that paper) shows \civbc\
emission FWHM$\la$5000 \kms\ consistent with the width of \civ\  emission in unabsorbed sources.

%\item BAL sources can also show mini-BALs that are either detached
%from (the \civ\ emission profile and/or the \civ\ BAL) or lie
%within the BAL trough. PG 1700+518 can be cited as our best
%example of a classic BAL while PG 2112+059 is the best example of
%a BAL with apparent mini-BALs within the BAL trough. The latter
%contains no attached emission line mini-BALs while the former
%shows a strong attached mini-BAL in the emission profile. PG
%1001+054 is an example of a reasonably detached mini-BAL
%superimposed on a shallow BAL absorption trough. These observations
%suggest that the BAL and mini-BAL phenomenologies are somehow related.
%This impression is reinforced by the discovery of associated mini-BAL
%features over the full range of blueshift velocity observed in BALs
%(e.g. up to 60,000km/s; Narayanan et al. 2004, Sabra et al. 2003).

\end{enumerate}

Individual sources  merit a few additional comments which are
given in Appendix A.

\subsection{BAL QSOs in the E1 Parameter Space \label{e1}}

BAL AGN with terminal wind velocities in the range $\approx$ 20000-30000 \kms\ are either outliers or are located
along the upper part of the E1 ``main sequence'' for Pop. A sources (Fig. \ref{fig:e1bal}) with a possibly different
distribution than the unabsorbed quasar distribution  \citep[as shown by][]{sulenticetal02} and also from that of the
mini-BAL AGN. The two sources that are located far from the ``main sequence'' and related model grid are hypothesized
to be very rare because few quasars of any kind are found in that region.(see \S \ref{outliers}).

The upper edge of the E1 ``main sequence'' has been modelled as the domain of  sources more highly inclined to our
line of sight (i.e., with the accretion disk seen more edge-on). The grid of \lm\ and orientation angle values
superposed on the optical E1 plane in Fig. \ref{fig:e1bal} was computed  for a fixed value of $M_{\rm BH}$$\approx
10^8$ \msol\ \citep[as in][]{marzianietal01}. This \mbh\ value is used because the average virial \mbh\ for our source
sample is $\log M \approx$\ 8.1 in solar units \citep[following][]{marzianietal03b,sulenticetal05}. There seem to be
no correlation with black hole mass. The dependence on mass in the E1 diagram is rather weak, and the dispersion of
$\log M \approx 0.6$\ is less than the mass difference needed to have effects comparable to the ones due to
orientation ($\Delta \log M \approx 1$, at least). However, sources with larger $M_{\rm BH}$ also tend to displace
upward \citep[i.e., toward larger FWHM(\hbbc);][]{zamanovmarziani02}. The location of the BAL QSOs in the optical E1
plane can be considered only as suggestive of larger inclination.

An orientation indicator appears to be emerging for the RQ Pop. A quasars as a low inclination blue outlier source
population \citep[showing a  significant \oiiiopt\ blueshift:][]{zamanovetal02,marzianietal03b}. Blueshifts of several
hundred \kms\ are observed in \oiiiopt\ for some NLSy1 (extreme E1 Pop. A sources), especially in sources with
unusually weak \oiiiopt\ \citep{zamanovetal02,marzianietal03b}. Blue outliers may involve the youngest type 1 sources
\citep{sandersetal88,zamanovetal02}. None of the 12 blue outliers with available UV data show BAL properties,
suggesting that BALs are not observed pole-on. The mini-BAL phenomenon appears to occur over a wider range of
inclination angles if our small sample can be taken as indicative. It is worth noting that NAL sources show no
occupation restriction in E1--they are found everywhere.

\subsubsection{E1 Outliers: Mrk 231 and IRAS 07598+6508}
\label{outliers}

We first point out that FWHM(\hbbc)$\approx$ FWHM(\feiiopt) within the uncertainty for all/most pop. A sources in our
sample of $\approx$ 300 low z AGN \citet{marzianietal03b}. The only two convincing exceptions are Mrk 231 and
IRAS07598+6508. Our E1 sample studies suggest that \hbbc\ in Pop. A sources can usually be modeled with a symmetric
Lorentzian function \citep{sulenticetal02}. Fig. \ref{fig:civhba} shows that Mrk 231, IRAS 07598, and perhaps a few
other sources, show a blue asymmetric \hb\ profile that could be modelled as a separate strongly blueshifted
component.  The presence/absence of this  extra component will affect FWHM measures for \hbbc\  (Fig.
\ref{fig:hbfits}). If one considers only the unshifted component with width comparable to FWHM(\feiiopt) the two
sources move into the region with FWHM(\hbbc)$\la$ 4000 \kms. These two sources are among the strongest \feiiopt\
emitters in our sample. If we consider only the \feii-like Lorentzian component of \hbbc\ in deriving the \rfe\
parameter we find a large increase to $\approx$5 and 3 for Mrk 231 and IRAS 07589+6508 respectively. If they are now
compared to the rest of the sample they will show the largest values of \rfe\ and will lie close to the upper FWHM
boundary for NLSy1s. The blueshifted line component roughly corresponds to the blueshifted \civ\ emission component
and it is not unreasonable to argue that part of the Balmer emission could arise in the postulated wind/outflow where
most of the \civ\ is produced. Alternatively, the blueshifted \hbbc\ residual could be associated to a nuclear
Starburst. A similar interpretation is possible for PG 0043+039 although the blueward asymmetry in the \hbbc\ profile
is below half maximum.

\subsection{A Terminal Velocity-Luminosity Correlation? \label{corr}}

Four out of five BAL QSOs with highest terminal velocities ($v_{\rm T} \la 15000$ \kms) are  Pop. A sources  which
have higher ${L}/{L_{\rm Edd}}$\ in general \citep{marzianietal03b}. Indeed, in our sample we find evidence for a
correlation between terminal wind velocity and absolute $V$\ magnitude (Fig. \ref{fig:corr}). This is expected if the
Eddington ratio is governing the wind/outflow as, for example, in the case of radiation pressure driven winds
\citep{laorbrandt02}.

The highest terminal velocities are likely possible only for Pop. A objects where we see the blueshifted \civ\
emission component. However the conclusion that  BAL QSOs are exclusive to Pop. A is challenged by the discovery of
radio-loud BAL QSOs in the FIRST survey \citep{beckeretal00} as well as by the presence in our sample of the Pop. B
source PKS 1004+13 (FWHM(\hbbc)$\approx$7000 \kms) that was identified as the first radio-loud BAL QSO relatively
recently \citep{willsetal99}.

The terminal velocity, in the case of a pressure driven wind, is $v_{\rm T} \propto \sqrt{L/L_{\rm Edd}}$. Considering
the classical BAL QSOs in our sample, we have

$$ v_{\rm T} \approx \frac{{\rm FWHM(CIV)}}{2} \sqrt{f_{\rm M} \frac{L}{L_{\rm Edd}}}$$

\noindent \citep[c.f.][]{laorbrandt02}. If we assume that the force multiplier $f_{\rm M}$\ is a function of
ionization parameter $U$\ (defined as the ratio of the photon density to electron density), well approximated by
$f_{\rm M} \approx 10^{2.55} U^{-\frac{1}{2}}$\ \citep{aravetal94}, we can write the terminal velocity as $$ v_{\rm T}
\approx 10^4 {\rm ~FWHM(CIV)}_{1000} U^{-\frac{1}{4}} (\frac{L}{L_{\rm Edd}})^\frac{1}{2}~ {\rm km s}^{-1},$$ where
FWHM(\civ) is in units of 1000 \kms. FWHM(\civ)/2 can be taken as a rough estimate of the initial injection velocity
in the case of a radial flow. If we assume: (1) that the most extreme pop. A sources radiate close to the Eddington
limit while Pop. B sources typically have ${L}/{L_{\rm Edd}} \approx 0.1$, and (2) that a proper FWHM(\civ) value is
4000 \kms\ and 6000 \kms\ for Pop. A and B respectively, we find $v_{\rm T} \sim 40000 U^{-1/4}$ \kms\ (Pop. A, $U \ga
0.01$) $> 20000 U^{-1/4}$ \kms\ (Pop. B, $U \la 0.5$).  Simple radiation-driven wind considerations are therefore
consistent with the existence of a Pop. B BAL source and are predicting systematically lower $v_{\rm T}$, as implied
by the observations. The terminal velocity in RL BAL QSOs may also be systematically lower than in RQ BAL QSOs
\citep{beckeretal00} as is true for our single example. Therefore, a Pop. B source does not necessarily undermine the
idea of a critical Eddington ratio accounting for the Pop. A-B dichotomy.

At this point in time  the evidence suggests that  RL or RQ Pop. B BAL sources are rare.  The NLSy1-like properties
(FWHM(\hbbc)$\approx$ 2100 \kms\ and strong \feiiopt) of J1556+3517 motivated our skepticism about the RL designation
for this source \citep{beckeretal97}. It is now known that it probably suffers from strong optical extinction
\citep{najitaetal00} and is RQ. This warrants caution about other claimed RL BALs especially if they show Pop. A
characteristics \citep[see][]{brothertonetal98,sulenticetal03}. PKS 1004+13 is a genuine RL pop B source, and unless
many more are found, one should consider it as a rare RL BAL source.

\section{Discussion}

\subsection{A Simple Model for Interpretation of the \civ\ Profiles \label{model}}

\begin{figure*}
\begin{center}
\includegraphics[width=14cm, height=7.6cm, angle=0]{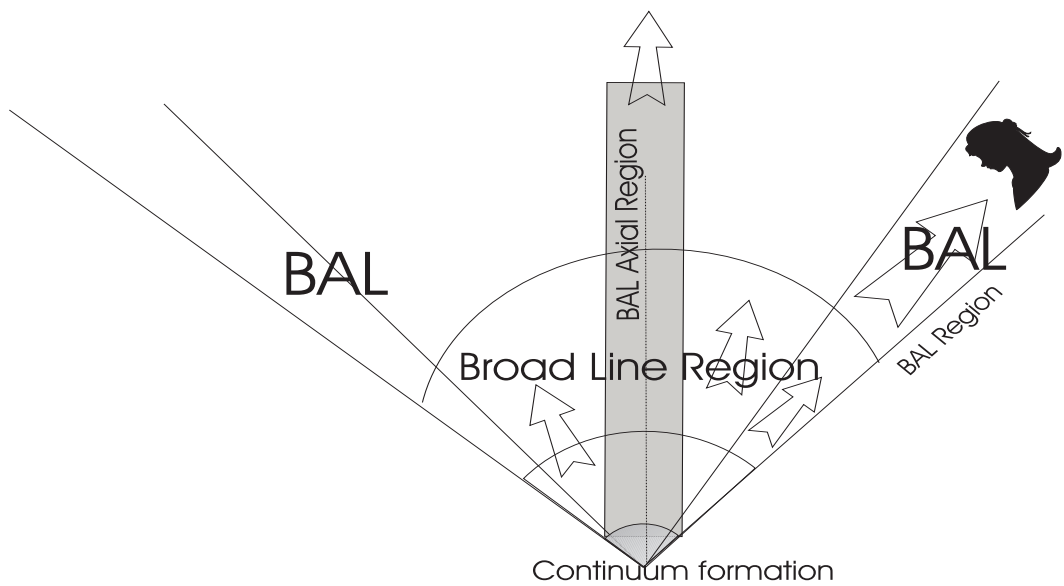}
\caption[]{Cartoon depicting the basic structure needed to account
for the BAL profiles discussed in \S \ref{model}. The observer
looking at the edge of a cone of opening angle $\approx$
100$^\circ$, sees a fully blueshifted emission component, a BAL
and a second, narrower BAL originating in an axial sheet of gas
participating in the general BLR outflow. See text for details.
\label{fig:model}}
\end{center}
\end{figure*}

Our results, coupled with the main constraints set by much earlier work \citep[see reviews
by][]{turnshek86,crenshawetal03}, lead to a simple geometrical and kinematical model (Fig. \ref{fig:model}). We
ascribe the high-ionization line emitting BLR to a wind as is now widely accepted. An immediate inference from {\em a
fully blueshifted} emission component involves a radial flow ($v/v_{\rm rot} \gg 1$) with an opening angle $\la
90^\circ$\ if BALQSOs are those QSOs viewed near the maximum possible orientation ($i \sim  45^\circ$). The BAL region
might extend in a conical corona of divergence angle $\la 10^\circ$ (which would include our line of sight; Fig.
\ref{fig:model}). The absorption profiles are furthermore consistent with a  radiation driven wind launched at {\em
roughly} the outer radius of the BLR because the emission component is only marginally ``eaten away" by the BAL. This
appears to be the case for the sources in this paper as well as for many BAL QSOs in high-$z$\ samples
\citep[e.g.,][]{koristaetal93}.

While this scheme is fully consistent with a ``standard BAL QSO model" that has emerged over the course of more than
15 years of research, the observation of secondary absorption components may require an additional element. We suggest
that whenever a secondary absorption is present we may be observing an axial region covering all of the
continuum-emitting region and part of the BLR as well. This property is needed to explain the depth of the absorption
which implies $f_{\rm c} \ga 0.5$. This axial region may involve  a cylindrical sheet of absorbing gas with different
physical properties than the BAL region. However, it appears to share the BLR flow because the absorption is located
close to the center of the \civ\ emission component in all three cases presented here. The narrower absorption profile
may result from a restricted viewing angle $\approx$50$^\circ$. It is important to stress that a shell of absorbing
material which could provide an adequate $f_{\rm c}$\ is not viable in the geometrical context of our model because of
the relatively narrow absorption.  A flow spiralling outward i.e., with  a significant rotational component may also
produce the secondary absorption although it seems again difficult to explain the absorption depth.

The cylindrical sheet  may have a straightforward physical explanation if it is connected with axial flows. A
two-fluid model for relativistic outflows includes a fast relativistic beam surrounded by a slower, possibly thermal,
outflow with a mixing layer between the beam and the jet \citep[e.g.][]{lobanovroland04}. A black hole with specific
angular momentum significantly different from zero, or even approaching the maximally rotating case ($a/M
\approx0.998$), is required for driving relativistic, radio-emitting jets \citep{blandfordznajek77}. The most
straightforward evidence for BH with $a/M \ga 0$, apart from the existence of radio jets, comes from Fe K$\alpha$\
profile shape and variations in a RQ Pop. A source, MCG --06--30--15 \citep{sulenticetal98}, as well as from energy
considerations for rapidly-varying X-ray sources \citep{forsterhalpern96} which are again RQ Pop. A sources. In a
scenario where some NLSy1s and, at least, some other Pop. A sources are young/rejuvenated quasars
\citep{mathur00,sulenticetal00a}, they may have experienced one of the rare accretion events leading to a consistent
increase of the black hole angular momentum \citep[unlike mass, black hole angular momentum can reverse and decrease
through BH mergings and accretion;][]{gammieetal04,hughesblandford03}. Accretion events leading to refueling of the
central BH are expected to be driven by mergers or strong interactions between galaxies. These give rise to an
enhancement in FIR emission (observationally) thought to reflect enhanced star formation activity
\citep[physically;][and references therein]{canalizostockton02,liparietal04}. Three sources in our sample are
Ultra-Luminous IR sources ($\log L_{\rm FIR} \ga 12$\ in solar units). They are the two E1 outliers  as well as PG
1700+518.  It is interesting to note that, on the basis of the \lm\ dependence shown in Fig. \ref{fig:e1bal}, the
observed \rfe\ for Mrk 231 and IRAS 07598+6508 would require highly super-Eddington accretion unless the accreting
object is a Kerr black hole. The only radio loud source in our sample of BAL QSOs (PKS 1004+13) is one of the 3
sources with a deep secondary absorption. As a final speculation, we therefore suggest that the inner cylinder may be
observed only in sources whose central black hole has a significant spin.

The proposed axial sheet of gas is actually a variant of the model proposed by \citet{elvis00} to explain the
occurrence of BAL and NAL. Bent flow lines seen at large viewing angles, or along the flow, could give rise to NAL and
BAL respectively. This would require that the flow be observable at large inclinations  ($i \rightarrow 90^\circ$)
which is not supported by the properties of the \civ\ profiles in our sample.

\subsection{From Low-$z$\ to High-$z$\ BAL QSOs \label{highz}}

It is interesting to consider the \civ\ emission properties of the few high-$z$\ BAL QSOs for which a rest frame can
be determined. Table \ref{tab:highzballah}  lists all cases that could be found in the literature. Following the
source name and redshift we list: FWHM(\hbbc) (Col. 3), qualitative evaluations of the strengths of \feiiopt\ and
\oiiiopt\ respectively (Col. 4-5), as well as  radial velocity shift of the \civ\ emission component with respect to
the rest frame (Col. 6, references are given in the footnotes). FWHM(\hbbc) and $\Delta v_{\rm r}$\ are given in the
quasar rest frame. We see that \civbc\ is most often shifted by $\ga$1000 \kms and overall the properties are more
consistent with Pop. A than Pop. B sources. This provides some support for the assumption that our sources are low
redshift analogues of the larger high redshift BAL population, even if our small sample cannot sample  all of the rich
high-$z$\ BAL QSO phenomenology. High-$z$\ BAL QSOs studied by \citet{hartigbaldwin86}\ show large \siiii/\ciii,
strong \feiiuv, low W(\civ) in emission and large blueshifts with respect to \mgii. These are also typical properties
of Pop. A sources \citep{zamanovetal02,bachevetal04}.

\subsection{BAL QSOs and the General QSO Population}

The model sketched in Fig. \ref{fig:model} also helps us to understand the relationship between BAL QSOs and general
Pop. A of quasars. Blueshifted \civ\ emission profiles have been known for a long time \citep{gaskell82} and, at low
$z$, the most extreme examples are observed in prototype NLSy1 sources like I Zw 1. Systematic \civ\ blueshifts are
also observed in the larger domain of Population A sources that include NLSy1s \citep{sulenticetal00b}.  The HIL \civ\
properties (and corresponding LIL \hbbc\ properties) were interpreted  as arising in a HIL wind \citep[from a LIL
emitting accretion disk;][]{marzianietal96,dultzinhacyanetal00}. If a source is observed almost pole-on the outflow
gives rise to a fully blueshifted \civbc\ profile. At the same time we observe the narrowest \hbbc\ profiles implying
that the outflow has a major velocity component perpendicular to the disk \citep[for a fixed \lm;][]{marzianietal01}.
This is consistent with the properties of e.g., I Zw 1 or Ton 28 with respect to the generality of Pop. A sources. The
blueshift of \civbc\ is smaller if the wind system is observed at larger angles. As the inclination increases, a
rotational component from the disk may  further reduce the \civbc\ profile shift. However, since we still observe an
almost fully blueshifted \civbc\ in several BAL QSOs of our sample, it seems unlikely that the inclination and the
half-opening angle of the radial flow exceed $\approx 45^\circ + 10^\circ \approx ~55^\circ$.

One must also take into account that $v/v_{\rm rot}$\ is expected to be a function of \lm, and hence to increase with
\rfe \citep{marzianietal01}. This lowers the likelihood of observing a large \civbc\ blueshift as one goes from
population A3 spectral types to B following the definition of \citet{sulenticetal02}. BAL QSOs with \civ\ profiles
like the PG 1700+518  one are therefore most likely Pop. A sources observed at extreme Eddington ratio and at higher
inclination. Their pole-on counterparts may be sources like I Zw 1.

The absence of any significant absorption at the \civbc\ line centre may be telling us that there is no significant
spin in the central engines. If black hole angular momentum is different than 0 we can expect to see the ``central"
absorption observed in BAL QSOs at the blue edge of the \civbc\ emission profile as a deteched mini-BAL. Such features
may be rare and of modest equivalent width.

\section{Conclusion}

This work supports the standard model for BAL QSOs (i.e., a radiation-driven wind flowing out at the edge of the BLR),
which directly accounts for most of the absorption/emission line properties observed in our small low-$z$\ sample. The
E1 correlations allow us to ascribe the most extreme BAL phenomenology to large Eddington ratio sources, and at the
same time to account for the existence of BALs in sources radiating at relatively low Eddington ratios (\S \ref{e1},
\ref{corr}). In the context of the E1 scenario it also suggests an axial structure associated with vertical ejection
contributes to the absorption, a suggestion requiring further testing.

This paper emphasizes the importance of having full coverage of the rest frame spectrum from 1500 \AA\ to the Balmer
lines. In this case the data allow to  accurately determine the quasar rest frame and to use the E1 parameter space
plus a simple geometric model to interpret  new results. This approach is now becoming possible for intermediate
redshift quasars ($1 \la z \la 2$) where IR spectra preserve  access to the region of \hb.
\newpage

\begin{figure*}
\includegraphics[width=17.6cm, height=17.6cm, angle=0]{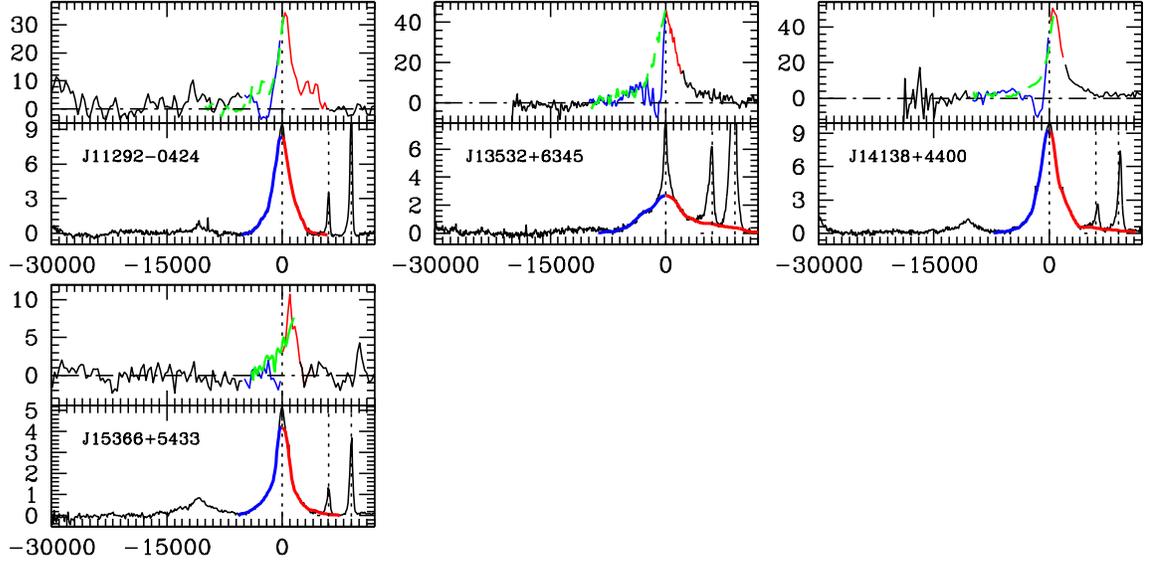}
\caption[]{Continuum and \feii\ subtracted profiles of \civ\ and
\hb\ for n=4 mini-BAL QSOs in our sample following Fig.
\ref{fig:civhba}. Dashed line (colored green) shows the assumed
unabsorbed profile used for absorption component measurements. In
the case of Mrk 486 ($\equiv$ J15366+5433), the thick green line
shows the HST \civ\ profile with no absorption. \label{fig:civhbb}
}
 \end{figure*}

\begin{figure*}
\begin{center}
\includegraphics[width=\columnwidth, angle=0]{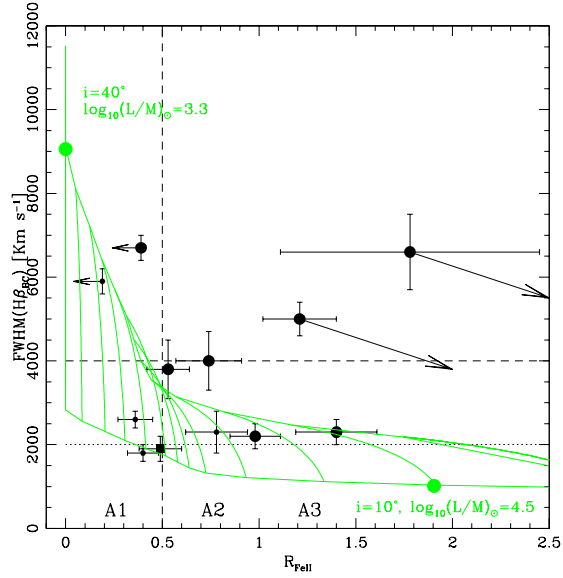}
\caption[]{The optical parameter plane of Eigenvector 1 with our
sample BAL and mini-BAL quasars indicated by large and small dots,
respectively.  Small and large dashed horizontal lines indicate
upper limits for NLSy1 and Pop. A sources. The green continuous
line traces the expected  location as a function Eddington ratio
and orientation, following \citet{marzianietal01}. The grid
represents the principal zone of low z source occupation as
modelled for an \mbh = $10^8$ \msol. The large, circular dots set
the scale of the grid indicating the expected position for $i=
40^\circ$\ and $\log$ \lm$\approx$ 3.3 (in solar units; upper
right) and for $i= 10^\circ$\ and $\log$ \lm$\approx$ 4.5 (lower
left). Lines tracing the orientation effect between minimum and
maximum $i$\ are incremented by $\Delta \log$\lm\ = 0.1. Arrows on
outlier sources (Mrk 231 and IRAS 07598+6508) indicate direction
of displacement after correction for a wind-related \hb\ component
not seen in most other low $z$\ quasars. The square identifies PG
1001+054 which has a highly uncertain absorption trough.
\label{fig:e1bal}}
\end{center}
\end{figure*}

\begin{figure*}
\begin{center}
\includegraphics[width=7.6cm, height=7.6cm, angle=0]{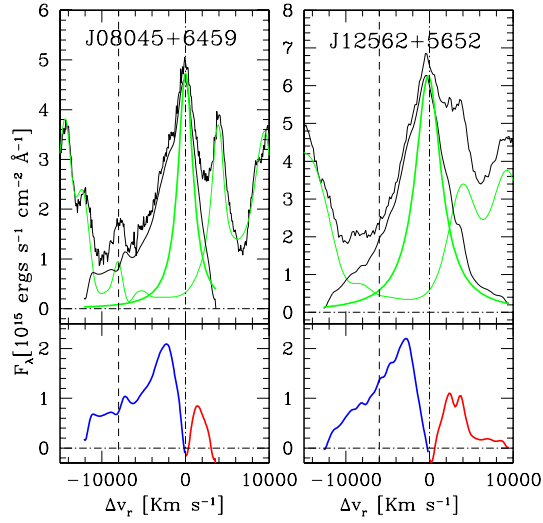}
\caption[]{Outlier sources IRAS 07598+6508 ($\equiv$ J08045+5459,
left) and Mrk 231 ($\equiv$ J12562+5652, right) \hbbc\ with
Lorentzian profile model of unshifted component most similar to
majority of sources (thick green line) using same width  \feiiopt\
lines ($\approx$ 2100 \kms; thin green lines). Original continuum
subtracted spectra shown as thin solid lines. Thick solid lines
show ``cleaned" \hbbc\ (after \feiiopt\ subtraction). Lower panel:
Residual components after subtraction of the Lorentzian profile
from \hbbc. Dashed line marks the radial velocity at which \civ\
absorption begins. Abscissa is radial velocity difference from
rest frame (\kms); ordinate is rest frame specific flux.
\label{fig:hbfits}}
\end{center}
\end{figure*}

\begin{figure*}
\begin{center}
\includegraphics[width=7.6cm, height=7.6cm, angle=0]{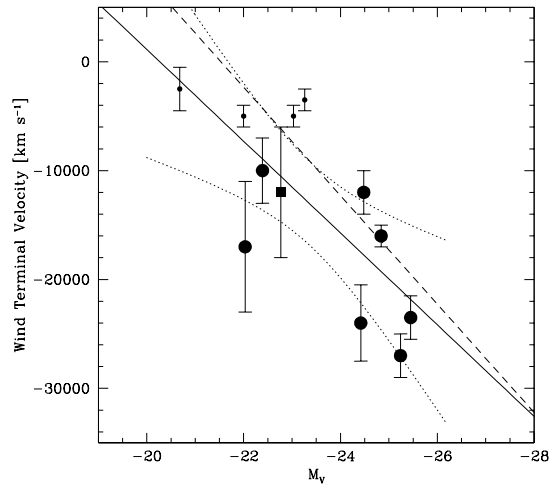}
\caption[]{Correlation between absolute $V$ magnitude $M_{\mathrm V}$\ and wind terminal velocity (in \kms). Symbols
used for data points have the same meaning of Fig. \ref{fig:e1bal}. The solid line is an unweighted least-square fit
made including all 12 data points. The dotted lines trace the $\pm$ 95\%\ confidence intervals. The dashed line is the
fit shown in Fig. 6 of \citep{laorbrandt02}, converted to the cosmology adopted in this paper. \label{fig:corr}}
\end{center}
\end{figure*}

\appendix{Properties of Individual Sources}
\begin{description}

\item PG 0043+039 $\equiv$ J00457+0410  -- shows a FIR excess but
only a marginal outlier in the E1 plane. The BALnicity index
reported goes to 0 because of the placement of the continuum,
which is likely to be more accurate in this study because of the
\feiiuv\ subtraction, even if the absorption trough remains
evident.

\item IRAS 07598+6508 $\equiv$ J08045+6459 -- This intriguing
object shows a FIR excess, and its location in the E1 plane is
peculiar. It is interesting to note that FWHM(\hbbc) $\approx$
5000 \kms\ $\gg$ FWHM(\feii) $\approx$ 2000 \kms. The good S/N
ratio and the strength of \feiiopt\ emission, along with the large
W(\hbbc) make this result especially striking (see \S
\ref{outliers} and Fig. \ref{fig:hbfits}).

\item PG 1001+054 $\equiv$ J10043+0513 -- Source PG 1001+054 has a
fairly uncertain absortion trough; could be a detached (or almost
detached) mini-BALs. The mini-BAL interpretation is supported by
the location of this source in the E1 diagram (Fig:
\ref{fig:e1bal}). There is no evidence of spectral variability
from the three spectra employed in this study.

\item PG 1004+130 $\equiv$ J10074+1248 -- The brightest radio loud
BAL QSO, recognized  by \citet{willsetal99}. The \civ\ absorption
and emission properties resemble those of the Pop. A BAL QSO with
BALnicity index $\gg$ 0, although this object is a lobe-dominated
radio loud quasar, with  FWHM(\hbbc)$\approx$ 7000 \kms.

\item PG 1126-041 $\equiv$ J11292$-$0424 -- A bona-fide mini
BAL-QSO.

\item Mrk 231 $\equiv$ J12562+5652  -- HST/FOS observations with
grism G190H produced only noisy spectra due to the strong internal
extinction. Mrk 231 is well known for the prototypical ULIRG and
mixed/Starburst AGN properties, and the SED shows an impressive
FIR excess over the typical RQ SED. The useful spectra were taken
with the low dispersion grism. The BALnicity index for \civ\ is
close to 0 \kms; however, this should be viewed with some care due
to the heavy internal reddening ($A_{\rm V} \approx 2$;
\citet{liparietal94}). \feii\ emission is not well reproduced by
the template, but this is more likely to  be due to the heavy
internal reddening rather than intrinsic differences with the I Zw
1 template. Since we modelled the observed \feiiq\ blend
(uncorrected for internal reddening), the actual \rfe\ could be
somewhat higher  than reported (but not dramatically so).

\item PG 1351+640 $\equiv$ J13532+6345   and PG 1411+441 $\equiv$
J14138+4400 -- They are mini-BAL QSOs, with the absorption eating
up the \civ\ line core at \dvr\ $\approx -$5000 \kms (less than
the emission blue wing zero intensity width). W(\civ) has been
estimated mirroring the red side of the profile, agrees very well
with the remaining emission blue wing at \dvr $\la$ --5000 \kms\
(\S \ref{meas}).

%\item PG 1552+085. The shallow absorption in \civ\ is confirmed by
%the \lya\ profile, albeit the low S/N of the spectrum makes the
%estimate of the wind terminal velocity and of W(\civ) highly
%uncertain.

\item Mrk 486 $\equiv$   J15366+5433   -- Mrk 486 shows a similar
absorption pattern as in PG 1351+640 and PG 1411+441 in a 1984 IUE
spectrum which is apparently no longer present in more recent
(1993) HST/FOS observations (see Fig. \ref{fig:civhbb} where the
profiles observed at different epochs are superimposed). However,
the relatively poor S/N of the HST spectrum makes it difficult to
ascertain whether the change in the mini-BAL absortion feature is
real.

\item PG1552+085 $\equiv$  J15547+0822  -- This source shows a
BALnicity index $\approx$0 \kms. We include this source among the
classical BAL QSOs because of its wide (albeit somewhat shallow)
absorption trough. An IUE spectrum has been published by
\citet{turnsheketal97}, and retrieved by us. The new STIS spectrum
confirms the BAL nature of the source, even if the feature, for
\civ\ is shallower and affected by noise since \civ\ is close to
the edge of the STIS covered spectral range.

\item PG 1700+518 $\equiv$  J17014+5149  -- PG 1700+518 is the
prototype low redshift BAL QSO. The UV spectral properties are
remarkably similar to the ones of Mrk 231, PG 0043+054, and IRAS
0759+6459.

\item PG 2112+059 $\equiv$  J21148+0607 -- This objects has a
\civ\ emission profile similar to the one of I Zw 1, and we
observe a fully detached absorption  trough. Optical lines are
significantly broader than in NLSy1s, but the object still lies
within the Pop. A area.

\end{description}

\begin{acknowledgements}
The authors acknowledge  support from the Italian Ministry of
University and Scientific and Technological Research (MURST)
through grant  and Cofin 00$-$02$-$004. D. D.-H. acknowledges
support from grant IN100703 PAPIIT UNAM. We thank M.H. Ulrich for
permission to use the ESO PG 0043+039 spectrum. This research has
made use of the NASA/IPAC Extragalactic Database (NED) which is
operated by the Jet Propulsion Laboratory, California Institute of
Technology, under contract with the National Aeronautics and Space
Administration.
\end{acknowledgements}

\newpage
%\bibitem{} Hall, P.~B., et al.\ 2002, ApJS, 141, 267
%\bibitem{} Hamann, F., Sabra, B., Junkkarinen, V., Cohen, R., \& Shields, G.\ 2002, X-ray Spectroscopy of AGN with Chandra and XMM-Newton, 121
%\bibitem{} Hamann, F.\ 1998, ApJ, 500, 798
%\bibitem{} Hamann, F.\ 1997, ApJS, 109, 279
%\bibitem{} Green, P.~J., Aldcroft, T.~L., Mathur, S., Wilkes, B.~J., \& Elvis, M.\ 2001, ApJ, 558, 109
%\bibitem{} Gaskell, C.~M., Brandt, W.~N., Dietrich, M., Dultzin-Hacyan, D., \& Eracleous, M.\ 1999, ASP Conf.~Ser.~175: Structure and Kinematics of Quasar Broad Line Regions,
%\bibitem{} Murray, N.~\& Chiang, J.\ 1995, ApJL, 454, L105
%\bibitem{} Piconcelli, E.,  Jimenez-Bailon E.,  Guainazzi M.,  Schartel N., Rodriguez-Pascual P.M.,  Santos-Lleo,M., 2004 astro-ph/0411051
%\bibitem{} Sabra, B.~M.~\& Hamann, F.\ 2001, ApJ, 563, 555
%\bibitem[Sulentic, Calvani, \& Marziani(2001)]{2001Msngr.104...25S} Sulentic, J.~W., Calvani, M., \& Marziani, P.\ 2001, The Messenger, 104, 25
%\bibitem{} Tanaka, Y.\ 1995, Seventeeth Texas Symposium on Relativistic Astrophysics and Cosmology, 759, 206

%\bibitem[Kr1997]{Kr97} Krabbe A., Colina L. and Thatte N. and Kroker A., 1997, ApJ 476, 98

\newpage

\begin{table*}
\begin{center}
\caption{Object Identification \& Optical Observations }
\label{tab:optobs}
    \begin{tabular}{lllllllll}
    \hline  \hline
    \noalign{\smallskip}
     IAU name & Object name &  \multicolumn{1}{c}{$m_{\mathrm V}$} & \multicolumn{1}{c}{$z$}   & \multicolumn{1}{c}{$M_{\mathrm V}$}
     & \multicolumn{1}{c}{Date} & \multicolumn{1}{c}{UT} & \multicolumn{1}{c}{ET} &\multicolumn{1}{c}{Obs.}\\
     \multicolumn{1}{c}{(1)}      & \multicolumn{1}{c}{(2)}         &    \multicolumn{1}{c}{(3)}
     & \multicolumn{1}{c}{(4)}    & \multicolumn{1}{c}{(5)} &  \multicolumn{1}{c}{(6)} & \multicolumn{1}{c}{(7)}
     & \multicolumn{1}{c}{(8)} & \multicolumn{1}{c}{(9)}\\
     \hline
     \noalign{\smallskip}
     J00457+0410   & PG 0043+039     &  16.0  & 0.3850 $\pm$ 0.0010     & -24.84& 09/10/1994 &  02:27 & 6000 & ESO \\
     J08045+6459   & IRAS 07598+6508 &  14.31 & 0.1488$^a$ $\pm$ 0.0001 & -24.42& 22/02/1991 &  02:18 & 2500 & KPNO \\
     J10043+0513   & PG 1001+054     &  16.21 & 0.1610 $\pm$ 0.0010     & -22.77& 19/02/1991 &  06:52 & 3600 & KPNO \\
     J10074+1248   & PG 1004+130     &  15.24 & 0.240$^b$  $\pm$ 0.001  & -24.57& 21/04/1990 &  05:10 & 3600 & BG   \\
     J11292$-$0424 & PG 1126-041     &  14.69 & 0.0560 $\pm$ 0.0001     & -22.00& 19/02/1991 &  08:16 & 1800 & KPNO \\
     J12562+5652   & Mrk 231         &  13.60 & 0.0412$^c$ $\pm$ 0.0001 & -22.39& 09/12/1994 &  09:14 & 3000 & SPM  \\
     J13532+6345   & PG 1351+640     &  14.3  & 0.0822$^d$$\pm$ 0.0001  & -23.26& 19/02/1991 &  10:38 & 1800 & KPNO\\
     J14138+4400   & PG 1411+442     &  14.65 & 0.0896$^d$ $\pm$ 0.0005 & -23.03& 19/02/1991 &  12:31 & 1800 & KPNO\\
     J15366+5433   & Mrk 486         &  14.8  & 0.0389$^e$ $\pm$ 0.0001 & -20.68& 01/07/1995 &  05:50 & 2400 & SPM \\
     J15547+0822   & PG 1552+085     &  16.3  & 0.119$^f$  $\pm$  0.001 & -22.03& 17/02/1990 &  12:17 & 2400 & BG  \\
     J17014+5149   & PG 1700+518     &  14.98 & 0.2899$\pm$ 0.0007      & -25.24& 30/04/1995 &  00:13 & 3600 & CA  \\
     J17014+5149   &                 &        &                         &       & 28/06/1995 &  07:27 & 3600 & SPM \\
     J21148+0607   & PG 2112+059     &  15.8  & 0.46005$\pm$  0.00010   & -25.45& 18/10/1990 &  01:43 & 2400 & KPNO\\     J21148+0607   &                 &        &                   &      & 29/06/1995 &  08:29 & 3600 & SPM \\
    \noalign{\smallskip}
    \hline \hline
    \end{tabular}
  \end{center}
  %\begin{list}{}{}
$^a$ Solomon et al. (1997) HI 21 cm;  optical determination
suggests larger uncertainty (shifted by -300 \kms). $^b$ Wills et
al. 1999.    $^c$ Average value of \citet{falcoetal99} and
\citet{devaucouleursetal91} (RC3). $^d$ \citet{marzianietal96}.
$^e$ SIMBAD. $^f$ \citet{schmidtgreen83}.
 %\end{list}
 \end{table*}

\newpage

 \begin{table*}
 \begin{center}
 \caption {LOG of UV   Archive Observations}
 \label{tab:uvobs}
    \begin{tabular}{llllcccl}
    \hline\hline
    \noalign{\smallskip}
     IAU name  &&  \multicolumn{6}{c}{UV Obs.} \\ \cline{3-8}
      &&Camera & Grating & Date & UT & ET (s) & Data Set \\

     \multicolumn{1}{c}{(1)}       && \multicolumn{1}{c}{(2)}     & \multicolumn{1}{c}{(3)}  &
     \multicolumn{1}{c}{(4)}& \multicolumn{1}{c}{(5)}    & \multicolumn{1}{c}{(6)}   & \multicolumn{1}{c}{(7)}
      \\
     \hline
     \noalign{\smallskip}
     J00457+0410  && HST/FOS  & G190H/BL      & 28/10/1991 & 04:36 &  1315 & Y0RV0204T \\
     J08045+6459  && HST/STIS & G140L         & 10/12/1998 & 16:37 &  2570 & O4YY18010 \\
                  && IUE/SWP  & Low disp.     & 28/09/1989 & 03:38 & 12840 & SWP37199  \\
                  && IUE/SWP  & Low disp.     & 17/12/1990 & 18:54 & 13200 & SWP40376  \\
     J10043+0513  && HST/FOS  & G190H/RD      & 16/01/1997 & 00:03 &  1260 & Y38O0208T \\
                  && HST/FOS  & G190H/RD      & 26/12/1996 & 17:27 &   530 & Y33S0304T  \\
     J10074+1248  && IUE/SWP  & Low disp.     & 12/01/1986 & 08:30 & 19000 & SWP27517  \\
                  && HST/STIS     & G140M     & 23/03/2003 & 03:23 & 2899 & O65EW03020\\
     J11292$-$0424  && IUE/SWP  & Low disp.   & 05/01/1985 & 12:05 &  9900 & SWP53285  \\
                  && IUE/SWP  & Low disp.     & 01/06/1992 & 07:31 &  5400 & SWP44822  \\
                 & & IUE/SWP  & Low disp.     & 01/06/1992 & 12:19 &  6600 & SWP44823  \\
     J12562+5652  && HST/FOS  & G150L/BL      & 21/11/1996 & 08:40 &   770 & Y3GS0404T  \\
     J13532+6345  && HST/FOS  & G190H/BL      & 05/09/1991 & 10:41 &  1500 & Y0P80306T \\
     J14138+4400 & & HST/STIS & G230L         & 12/02/2001 & 01:23 &  1200 & O65617010 \\
     J15366+5433  && HST/FOS  & G190H/BL      & 15/11/1993 & 02:56 &  1440 & Y1EL0302T \\
                  && IUE/SWP  & Low disp.     & 07/05/1984 & 00:23 & 51600 & SWP22932  \\
     J15547+0822  && IUE/SWP  & Low disp.     & 28/04/1986 & 13:40 & 10200 & SWP28237  \\
                  && HST/STIS & G230L         & 05/07/2003 & 07:23 &  900  & O6MZ4I010 \\
     J17014+5149  && HST/STIS & G230L         & 30/01/2001 & 16:27 &  1200 & O65619010 \\
     J21148+0607 & & HST/FOS  & G190H/RD      & 19/09/1992 & 14:27 &  2372 & Y10G0204T \\
                  && HST/FOS  & G270H/RD      & 19/09/1992 & 16:07 &  1690 & Y10G0205T \\
    \noalign{\smallskip}
    \hline\hline
  \end{tabular}
  \end{center}
 \end{table*}

\newpage

 \begin{table*}
 \begin{center}
  \caption {\hbbc\ and \feii }
  \label{tab:hb}
%   \scriptsize
    \begin{tabular}{lcccc}
     \hline\hline
     \noalign{\smallskip}
     \noalign{\smallskip}
     IAU name & W(\hbbc) &  \rfe  &
     FWHM(\hbbc) & FWHM(\feiiq) \\
     \noalign{\smallskip}
      & [\AA]  &  & [\kms] & [\kms] \\
     \multicolumn{1}{c}{(1)}         & (2)   & (3)           & (4)   & (5)  \\
     \hline
J00457+0410    &   101 $\pm$ 11   & 0.74 $\pm$   0.17  &  4000 $\pm$   700 & 4300   $\pm$ 1000 \\
J08045+6459    &   77  $\pm$ 7    & 1.21 $\pm$   0.19  &  5000 $\pm$   400 & 2100   $\pm$  300 \\
J10043+0513    &   93  $\pm$ 5    & 0.49 $\pm$   0.11  &  1900 $\pm$   300 & 1850   $\pm$  300 \\
J10074+1248    &   37  $\pm$ 4    &  $\la$0.39         &  6700 $\pm$   300 &       \nodata     \\
J11292-0424    &   82  $\pm$ 8    & 0.78  $\pm$  0.16  &  2300 $\pm$   500 & 3100   $\pm$  800 \\
J12562+5652    &   45 $\pm$  15   & 1.78  $\pm$  0.67  &  6600 $\pm$   900 &   $\approx$ 3000  \\
J13532+6345    &   34 $\pm$  4    &   $\la$0.19        &  5900 $\pm$   300 &       \nodata     \\
J14138+4400    &   86 $\pm$  4    & 0.36   $\pm$ 0.09  &  2600 $\pm$   200 & 2600   $\pm$  500 \\
J15366+5433    &   84 $\pm$  4    & 0.40   $\pm$ 0.08  &  1800 $\pm$   200 & 1850   $\pm$  500 \\
J15547+0822    &   54 $\pm$  5    & 1.40   $\pm$ 0.21  &  2300 $\pm$   300 &       \nodata     \\
J17014+5149    &   55 $\pm$  5    & 0.98   $\pm$ 0.13  &  2200 $\pm$   300 & 1700   $\pm$  400 \\
J21148+0607    &   118 $\pm$ 11   & 0.53  $\pm$  0.11  &  3800 $\pm$   700 & 4050   $\pm$ 1000 \\
     \noalign{\smallskip}
     \hline
   \end{tabular}
 \end{center}
 \end{table*}

\newpage

 \begin{table*}
 \begin{center}
  \caption {\civ}
  \label{tab:civ}
%   \scriptsize
    \begin{tabular}{lccccccccc}
     \hline\hline
     \noalign{\smallskip}
     \noalign{\smallskip}
     IAU name & \multicolumn{3}{c}{Emission}&&
     \multicolumn{4}{c}{Absorption}\\  \cline{2-4} \cline{6-10}
     \noalign{\smallskip}
     &  $v_{\rm r}$ & W({\sc Civ}) & FWHM({\sc Civ}) && $v_{\rm r}$ & $v_{\rm r,T}$ & W({\sc Civ}) & FWHM({\sc Civ}) &  Baln.I.\\
      & [\kms] & [\AA]& [\kms] && [\kms] &[\kms] & [\AA]& [\kms]& [\kms]\\
     \multicolumn{1}{c}{(1)}         & (2)   & (3)           & (4)     && (5) & (6) & (7) & (8)&(9)    \\
     \hline
     \noalign{\smallskip}
     J00457+0410 & --1800   &  10  &  8000                                              && --11300  & --16000 & --11    & 7300                   & 0 \\
     J08045+6459 & --5000: & 33: &  6000:                                               && --13400  & --27000 & --40   & 9000   & 5300\\
                &                          &                  &                         && --3900   &         & --10 & 1160  &  0\\
     J10043+0513 & --1300                   & 52               &  3800                  && --8700   & --12000 & --5 & 3000    & 400 \\
     J10074+1248 & --3000:                  & 3$_{-2}^{+2}$    &  6000:                 && --6700   & --12000 & --12   &  5950                   & 1050 \\
                 &                         &                   &                        && --930    &         & --4&  1500                     &0  \\
     J11292$-$0424 & +900$_{-1000}^{+200}$ & 42:               &  2200:                 && --2400   & --5000  & --4.6&  1500     & 0 \\
     J12562+5652 & +0$_{-1000}^{+500}$     & 16               &  5900                   && --8300   & --10000 & --5  &  3400     & 50 \\
     J13532+6345 & +0                      & 31:              &  2400                   && --1200   & --3500  & --5.5  &  1800    & 0 \\
     J14138+4400 & +300$_{-1000}^{+200}$   &  36:              &  2200:                 && --1450   & --4000  & --7$_{-7}^{+2}$  &  1000$_{-500}^{+2000}$   & 0 \\
     J15366+5433 & +1000                   &  18              &  1500                   && --800    & --2500  & --4&  1500   & 150 \\
     J15547+0822 & --1000$_{-500}^{+500}$   & 60               &  3700                  && --11000  & --17000 & --6$_{-6}^{+2}$ &  5000:   & 500 \\
     J17014+5149 & --3100:                   & 19$_{-10}^{+10}$&  7000:                 && --15000  & --27000 & --73   &  14200  & 10300 \\
               &                           &                    &                       && --5000   &         &  --4.5   &   2200    & 0\\
     J21148+0607 & --550                    & 23   & 4600                               && --15600  & --26000 &  --22   &  9700    & 2500 \\
     \noalign{\smallskip}
     \hline
\multicolumn{10}{l}{A double colon mark indicates highly uncertain values.}\\
   \end{tabular}
 \end{center}
%$^a$: Indicative width at half maximum of emission feature which
%includes absorption component in the middle. It is highly
%uncertain especially for J10074+12

 \end{table*}

\newpage

\begin{table*}[!t]\centering
%  \newcommand{\DS}{\hspace{6\tabcolsep}} %% Expanded Space between
  %% some cols
%  \setlength{\tabnotewidth}{0.9\textwidth}
%  \setlength{\tabcolsep}{1.3\tabcolsep}
  \tablecols{6}
  \caption{High  $z$ BAL or mini-BAL QSOs with \civ\ Shift\tabnotemark{a}} \label{tab:highzballah}
  \begin{tabular}{llccrc}
    \toprule
    \multicolumn{1}{c}{Object}
    & \multicolumn{1}{c}{$z$}
    &\multicolumn{1}{c}{FWHM(\hb)}
&\multicolumn{1}{c}{\feiiopt}&\multicolumn{1}{c}{[O{\sc iii}]}
&\multicolumn{1}{c}{$\Delta v_{\rm r}$
(\civ)}\\
& & [\kms] & & &\multicolumn{1}{c}{[\kms]}   \\
\multicolumn{1}{c}{ (1)  }       & \multicolumn{1}{c}{(2)}   & (3)           & (4)     & \multicolumn{1}{c}{(5)} & (6)    \\
\midrule
0043+008(UM275)  &  2.146\tabnotemark{b}  &  4300:                 &    weak      &  mod.  &     0\tabnotemark{c} \\
                 &  2.1526\tabnotemark{d}  &                       &              &        & -500\tabnotemark{c}  \\  %
0226-038 (RL?)   &  2.073\tabnotemark{b}  &  2800:                 &    weak      & strong &   -900\tabnotemark{e}\\ %
0842+345(CSO203) &  2.163\tabnotemark{b}  &  8400:                 &      strong  &  weak  & -3000\tabnotemark{c} \\ %
1011+091         &  2.305\tabnotemark{b}  &  7500:                 &      strong  &weak    & -4000\tabnotemark{c} \\ %
1246-057         &  2.244\tabnotemark{f}  &  4500\tabnotemark{g}   &   strong     &  weak  &-2000\tabnotemark{c}  \\
                 &  2.243\tabnotemark{b}  &  5900:                 &   strong     &   weak & -2000\tabnotemark{c} \\
                 &  2.246\tabnotemark{h}  &                        &              &        &-2200\tabnotemark{c}  \\
1309-056         &  2.220\tabnotemark{b}  & 3200:                  &      strong  & weak   & -750\tabnotemark{c}  \\
                 &  2.239\tabnotemark{h}  &                        &              &        &-2500\tabnotemark{c}  \\
H1413+117        &  2.551\tabnotemark{f}  &  4000\tabnotemark{g}   &     strong   & strong & 0\tabnotemark{j}     \\
                 &  2.558\tabnotemark{i}  &                        &              &        &  -600\tabnotemark{j} \\
LBQS2212-1759    &  2.228\tabnotemark{b}  &  6100:                 &     weak     & strong & -600\tabnotemark{c}  \\
\bottomrule \multicolumn{6}{l}{$^a$ Sources for which a reliable
measurements of systemic $z$ and \civ\ emission component shift
exist.}\\
\multicolumn{6}{l}{$^b$  From \oiiiopt\ as reported in \citet{mcintoshetal99}.}\\
\multicolumn{6}{l}{$^c$  \citet{koristaetal93} UV/\civ\ redshifts.}\\
\multicolumn{6}{l}{$^d$  From SDSS.}\\
\multicolumn{6}{l}{$^e$  UV $z$\ from \citet{lanzettaetal87}}\\
\multicolumn{6}{l}{$^f$  \citet{hilletal93}.}\\
\multicolumn{6}{l}{$^g$  \citet{hilletal93} --  \ha.}\\
\multicolumn{6}{l}{$^h$  \citet{espeyetal89} --  \ha.}\\
\multicolumn{6}{l}{$^i$  \citet{barvainisetal97}  from   CO.}\\
\multicolumn{6}{l}{$^j$  \civ\ emission; \citet{angoninetal90}.}\\
\end{tabular}
\end{table*}
\hfill
\newpage

\newpage

%\tableofcontents
\end{document}